\documentclass[journal=jacs,manuscript=article,layout=twocolumn]{achemso}

\usepackage[version=3]{mhchem} % Formula subscripts using \ce{}
\usepackage{graphicx}
\usepackage{times}
\usepackage{caption}
\usepackage{hyperref}
\usepackage{xurl}
\usepackage{booktabs}

\author{Zachary T.P. Fried}
\email{zfried@mit.edu}
\affiliation[MIT]
{Department of Chemistry, Massachusetts Institute of Technology, Cambridge, MA 02139}
\author{Brett A. McGuire}
\email{brettmc@mit.edu}
\affiliation[MIT]
{Department of Chemistry, Massachusetts Institute of Technology, Cambridge, MA 02139}

\title{Automated Mixture Analysis via Structural Evaluation}

\abbreviations{IR,NMR,UV}
\keywords{American Chemical Society, \LaTeX}

\begin{document}

\begin{tocentry}

\includegraphics[width=\textwidth]{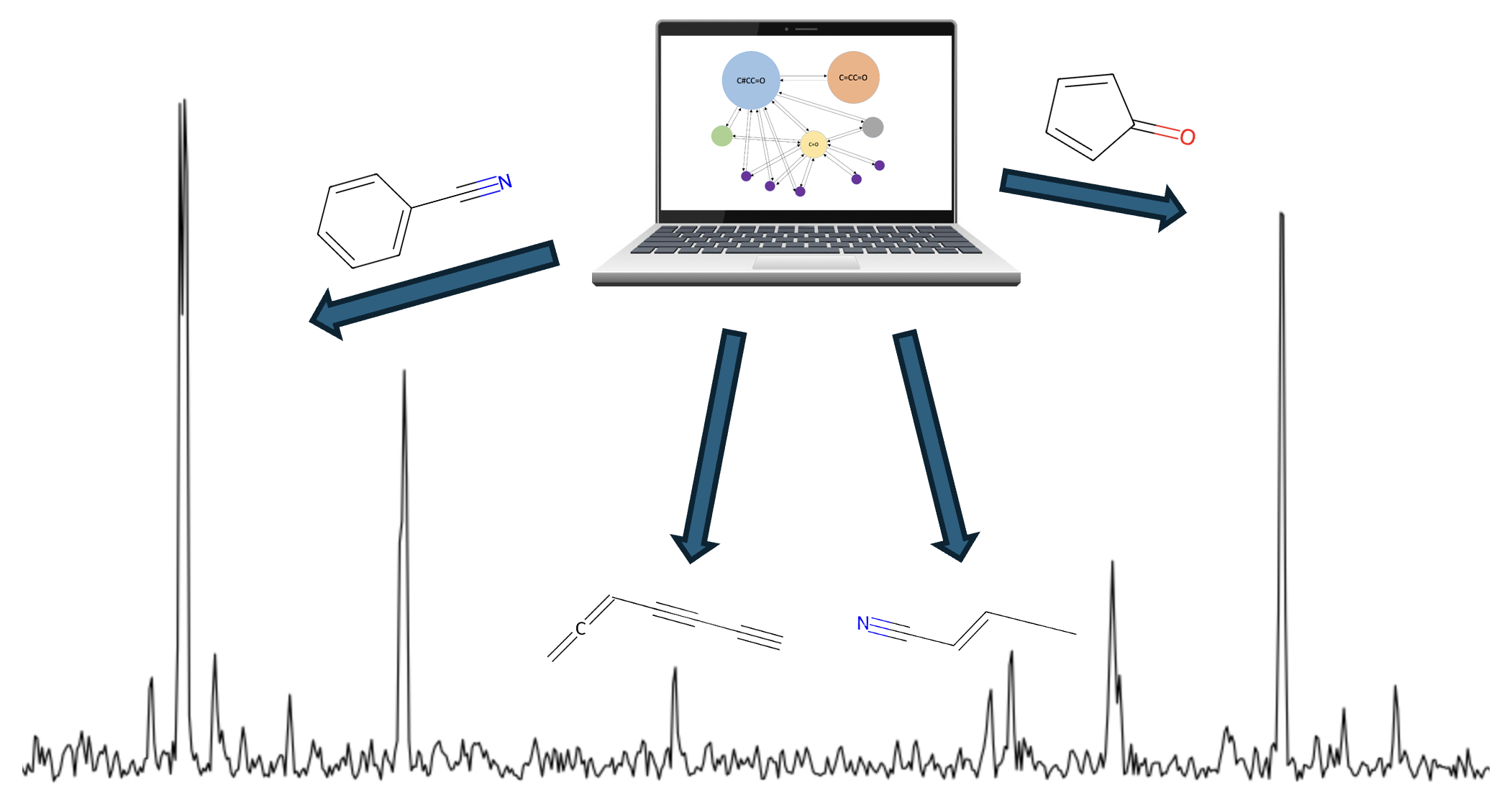}

\end{tocentry}

\begin{abstract}
  The determination of chemical mixture components is vital to a multitude of scientific fields. Oftentimes spectroscopic methods are employed to decipher the composition of these mixtures. However, the sheer density of spectral features present in spectroscopic databases can make unambiguous assignment to individual species challenging. Yet, components of a mixture are commonly chemically related due to environmental processes or shared precursor molecules. Therefore, analysis of the chemical relevance of a molecule is important when determining which species are present in a mixture. In this paper, we combine machine-learning molecular embedding methods with a graph-based ranking system to determine the likelihood of a molecule being present in a mixture based on the other known species and/or chemical priors. By incorporating this metric in a rotational spectroscopy mixture analysis algorithm, we demonstrate that the mixture components can be identified with extremely high accuracy ($\geq97\%$) in an efficient manner. 

\end{abstract}

\section{Introduction}
Chemical samples almost ubiquitously come in the form of complex mixtures as opposed to pure compounds. The determination of the components of such  mixtures is of central importance in applications ranging from pharmaceuticals and food sciences to environmental chemistry and astrochemistry. A multitude of analysis techniques are commonly used for the qualitative or quantitative characterization of these mixtures, including microwave, infrared, Raman, UV-Vis, X-ray, and NMR spectroscopies along with the coupling of liquid chromatography or gas chromatography with mass spectrometry (LC-MS and GC-MS) \cite{dumez22,fiehn16,masson10,ryder99,uv-vis,rot,xray}. The utility of each of these techniques heavily depends on the nature of the sample and the analysis.  Yet, in every case the signals generated and interpreted are uniquely molecule-specific.  That is to say, every signal carries molecular information that can be used to interpret and uniquely assign these features to a molecular carrier.  Doing so in an automated fashion is highly desirable both for fundamental research and for applications from industry to field measurements and monitoring.  Here, we describe a novel and technique-agnostic approach to achieve this goal.  We apply this approach to rotational spectroscopic analysis of chemical mixtures in a proof-of-concept demonstration.

Recent developments in machine-learning based chemical vector embedding models provide a straightforward way to mathematically calculate structural/chemical similarity \cite{wigh22}. We exploit these embeddings to assign each molecule a unique location in ``chemical space."  By doing so, the similarity between any two molecules can be quantitatively represented by the distance between those molecules in chemical space. Additionally, graph-based algorithms, such as Google's PageRank \cite{page_rank}, have been effectively used to explore relationships between objects in datasets.  We therefore construct a graph-based architecture, with connectivity determined by the distances between the chemical embeddings, to rank molecular candidates by their chemical relevance to a specific mixture. The resulting rankings are then incorporated into automated assignment algorithms for chemical mixtures, resulting in accurate and efficient assignments of observed signals, their carriers, and thus the mixture components.

One technique that has been fairly under-utilized with regards to mixture analysis is rotational spectroscopy. This method measures transitions in the rotational states of freely rotating molecules in the gas phase following interaction with radiation. The efficacy of this method with regards to mixture characterization stems from its high structural sensitivity and spectral resolution \cite{park16}. The spacing of the rotational transitions directly depends on the molecules' moment of inertia along three principal axes. Therefore, even structurally similar conformers, isomers, and isotopologues have completely distinctive rotational spectra since the mass distributions are unique. The highly resolved and oftentimes narrow spectral features that result from this method limits peak overlapping, theoretically making identification of individual carriers straightforward. 

The advent of broadband \cite{brown08} and semi-automated rotational spectroscopy techniques \cite{mst,amdor} has resulted in an explosion in the number of molecules characterized and their spectra catalogued in the literature and public databases. While rotational transition frequencies are entirely unique to each molecule, and uncertainties are often very small, the sheer number of catalogued transitions means that oftentimes several potentially viable molecular carriers have rotational transitions nearby any measured frequency in the microwave regime. Thus, when analyzing a mixture using rotational spectroscopy, considering only the closeness of frequency match is generally not sufficient for accurate characterization. That being said, in a large number of natural or experimentally produced mixtures, the mixture components are chemically or structurally related either via similar production pathways or environmental factors. Therefore, having a metric to gauge the chemical/structural likelihood of a molecule within a chemical mixture is vital for the accurate and automated assignment of mixture components from spectral features. 

By incorporating our new analysis methodology into an automated assignment algorithm for chemical mixtures using rotational spectroscopy, we demonstrate that this ranking system provides accurate assignment of the mixture components in an efficient manner. We also discuss the potential application of this method to additional spectroscopic techniques. 

\section{Methodology}
\subsection{Technique-Agnostic, General Approach}
The bulk of this paper is dedicated to describing the detailed application of this algorithm (titled ``Automated Mixture Analysis via Structural
Evaluation (AMASE)") to the analysis of mixtures using rotational spectroscopy.  AMASE is technique-agnostic, however, and so we first describe the overall workflow, highlighting where application-specific customizations are made before moving into our proof-of-concept experiments.

\begin{figure*}[tbh!]
\begin{center}
\includegraphics[width=\textwidth]{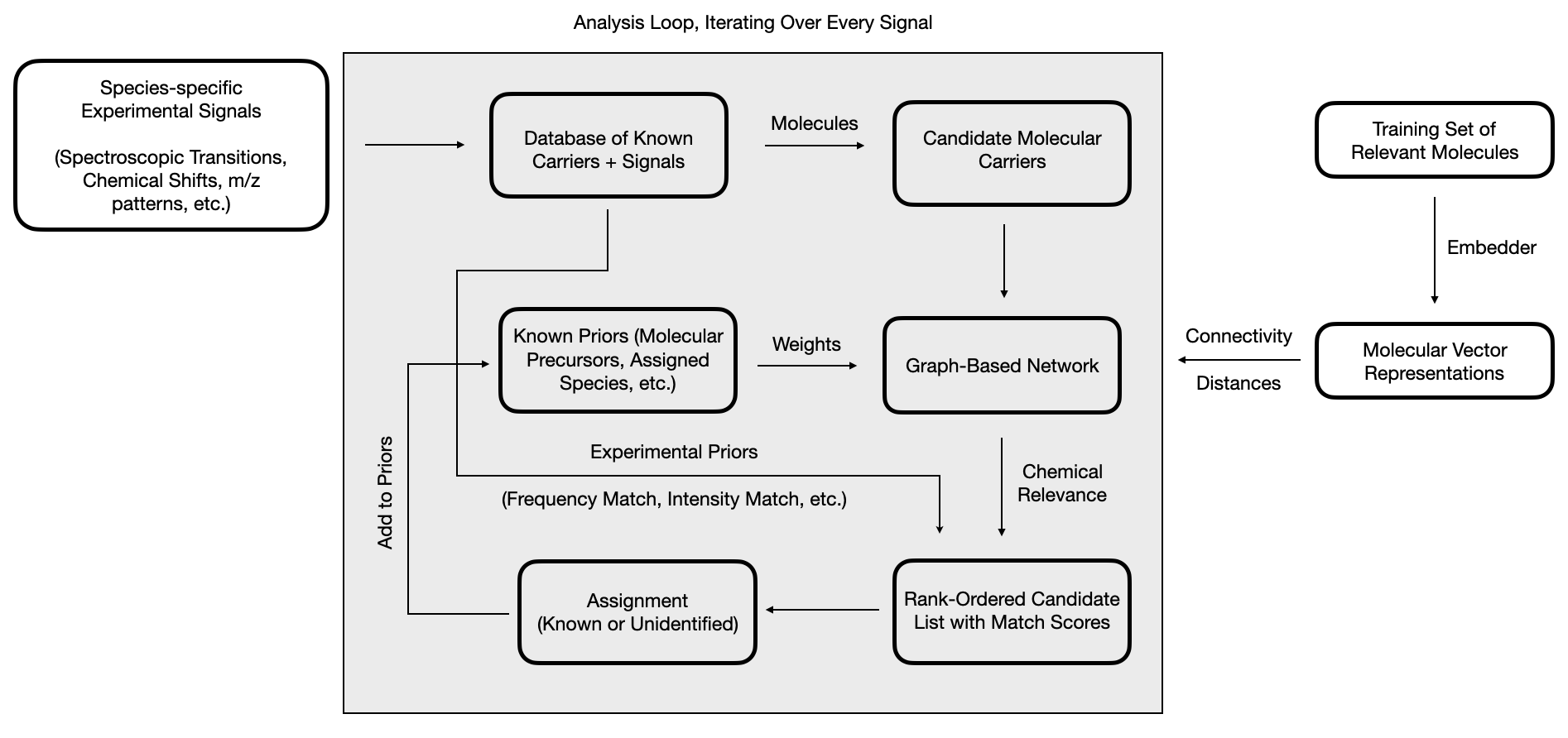}
\caption{General schematic of the automated assignment process.}\label{fig:general}
\end{center}
\end{figure*}

The general schematic for AMASE is shown in Figure~\ref{fig:general}.  AMASE works with any dataset that consists of distinct signals that arise from molecule-specific properties.  This can be an NMR spectrum consisting of chemical shifts, a mass spectrum consisting of $m/z$ ratios, a rotational/vibrational/electronic absorption or emission spectrum, and so forth.  AMASE interfaces with a database (online and/or local) consisting of the known signals of molecules of interest and then attempts to confidently assign the carrier of every observed experimental signal as either belonging to a specific molecule in the database or one that is currently unknown.

The experimental signals are sorted (often by either absolute strength or signal-to-noise ratio) and iterated through one at a time.  The database is queried for all signals from known molecules which fall reasonably close to the observed signal.  This then results in a list of candidate species and their associated candidate signal assignments.

Each of these candidate species and signals will eventually be assigned a ranked score by AMASE, indicating the confidence that this is the carrier of the experimental signal.  A variety of factors, each with their own weight, are considered.  For example, in spectroscopic investigations, contributions to the match can include how close the signal falls in frequency to the experimental line, the uncertainty of the catalog line, and whether the other transitions of the species which should be visible are present in the data. 

Additionally, AMASE employs a graph-based network to leverage chemical information and incorporate it into the matching criteria.  Before the graph can be constructed, a large training set of molecules (typically a few million) is gathered from various databases and used along with a molecular embedding algorithm to construct molecular feature vectors.  These feature vectors are numerical representations of molecules that encode learned structural and chemical information for each species in the context of the full training set.

A graph-based network is then constructed from a subset of the species (typically a few hundred thousand) that are likely to be chemically relevant to the mixture under study.  Each species is represented as a node in the graph, and the connectivity and structure of the graph is then informed by the relationships between the molecular feature vectors.  If any species are known to be present in the mixture \emph{a priori}, these are initialized with a large weight to that node in the graph to represent their importance.  Weights are then iteratively distributed throughout the graph, based on connectivity, such that species nearby those known to be present have increased weight as being more likely carriers of signals.

AMASE then compiles a final match score for each potential assignment based on the experiment-specific criteria outlined above combined with the likelihood of each species from the graph. It then generates a rank-ordered list of candidate signals and carriers, and assesses whether the top-ranked candidate passes a threshold to be confidently assigned. The current confident assignment thresholds have performed adequately for all of the mixtures tested so far. However, if needed, they can be easily adjusted in the source code.

If a new species is assigned to be present, it is added to the list of known molecular priors and the weights in the graph are updated to reflect the new importance of this node.  The entire process is then repeated for the next line in the assignment list.  After each assignment, all prior assignments are iteratively re-checked to ensure that the increasingly large pool of information gleaned from the process does not suggest a prior misassignment. In the end, AMASE produces for each signal either a confident assignment, a suggested assignment with potential alternatives, or a label of unidentified, indicating that the species is either not in the database containing known signal assignments or insufficient evidence exists to make an assignment.\\

\subsection{Algorithm Design for Rotational Spectroscopy}
\begin{figure*}[th!]
\begin{center}
\includegraphics[width=\textwidth]{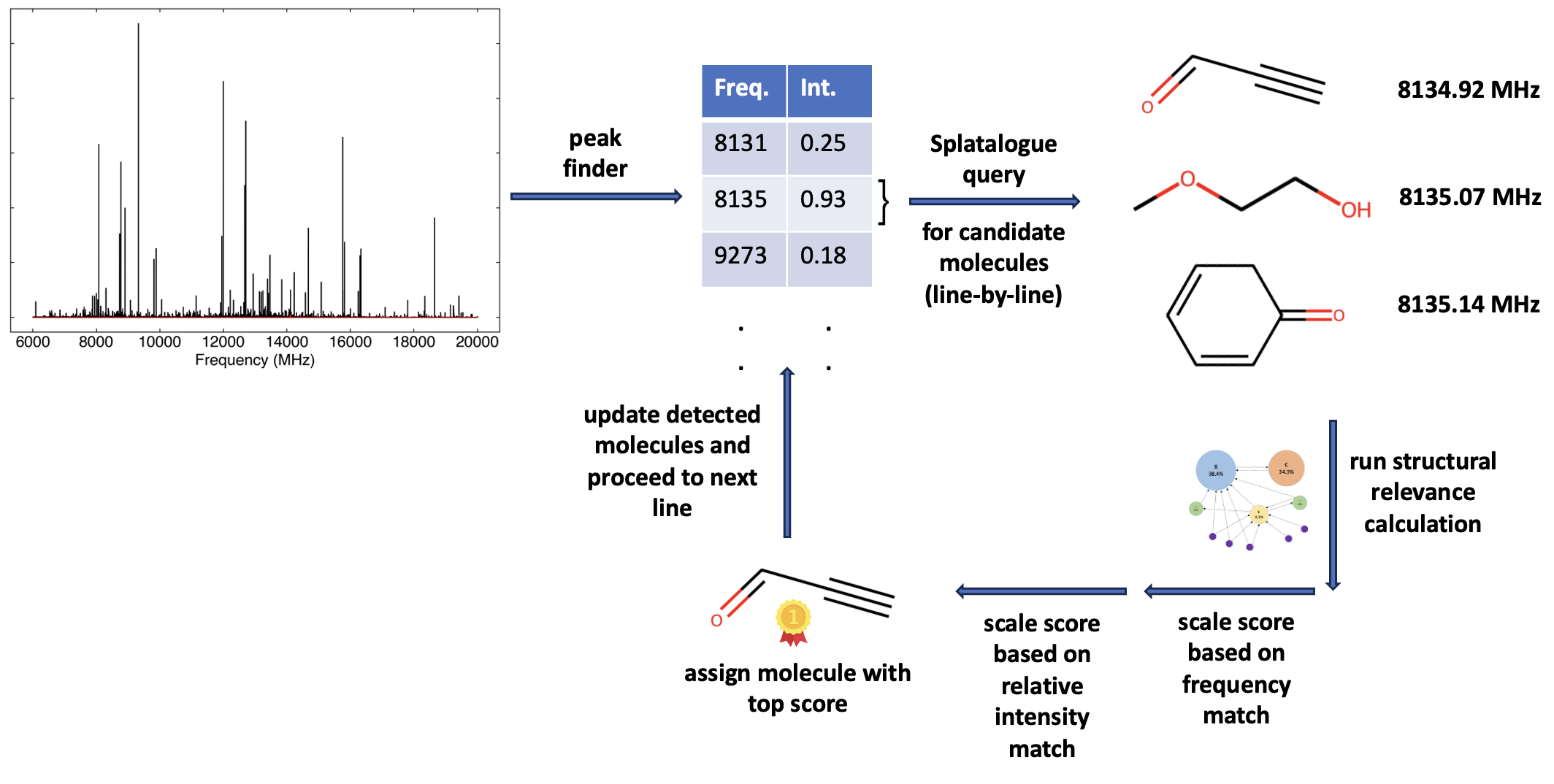}
\caption{Schematic of the line assignment process for a mixture studied with rotational spectroscopy. The pictured spectrum was collected by McCarthy et al. (2020) \cite{discharge}. The molecules in the mixture are discharge products of benzene and \ce{O2}.}\label{fig:schematic}
\end{center}
\end{figure*}

Now, we walk through this procedure, highlighting each step for the process of assigning the components of a mixture using rotational spectroscopy.  We discuss each decision point, the tunable hyperparameters used, and demonstrate the results on three different mixtures. A schematic of this experiment-specific process is depicted in Figure~\ref{fig:schematic}.

\begin{figure*}[th!]
\begin{center}
\includegraphics[width=\textwidth]{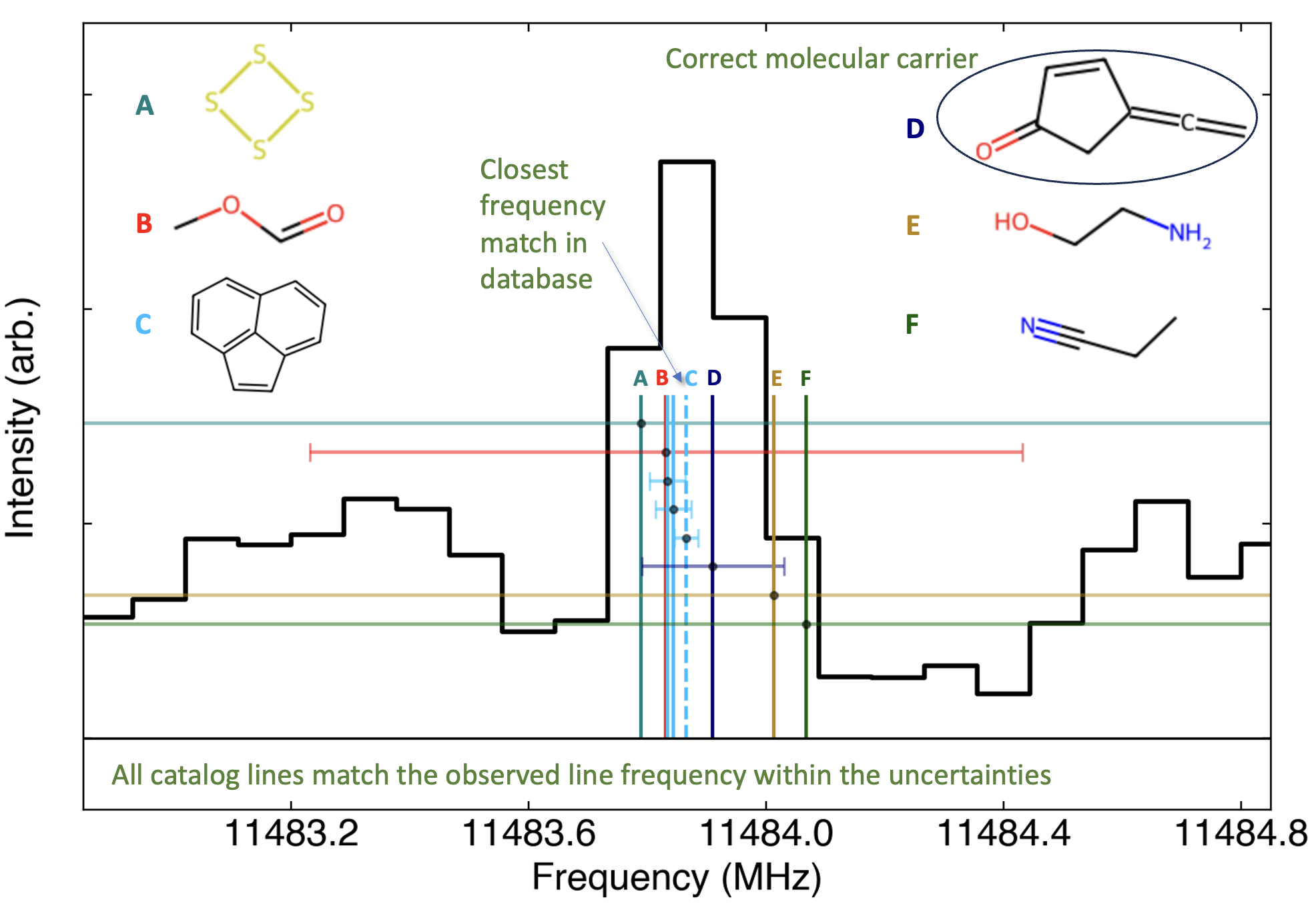}
\caption{Known rotational transitions within 0.3 MHz of a transition observed in the benzene/\ce{O2} discharge experiment conducted by McCarthy et al. (2020) \cite{discharge}. The dashed line is the transition with the closest frequency to the observed peak. This line, however, does not correspond to the correct molecular carrier. 4-ethenylidene-cyclopent-2-en-1-one (molecule D) is in fact the true carrier of this line in the mixture. The error bars are 10 times the statistical catalog uncertainties. This scaling factor was employed because it is well known that the statistical uncertainties listed in rotational spectroscopy databases like Splatalogue commonly underestimate the true values (shown in the work of Melosso et al. (2020) \cite{splat_unc}).}\label{fig:candidates}
\end{center}
\end{figure*}

When ranking molecular candidates for each transition in an observed rotational spectrum, the structural/chemical relevance of the molecule is considered along with the frequency and intensity match of the simulated spectrum for a given species. Figure~\ref{fig:candidates} illustrates the importance of considering more than just the frequency match when deciding the correct molecular carrier. This plot shows a spectral peak that appears in the benzene/\ce{O2} discharge dataset that is described later. Overlaid on top of it are eight known rotational transitions within 0.3 MHz of the center frequency corresponding to six unique molecules.  In this case, the catalog line with the closest frequency to the observed transition is actually not the correct molecular carrier. In fact, all of the transitions have error bars that overlap with the center frequency of the observed peak.

The first step in this algorithm is to determine the frequencies of the peaks in the mixture spectrum using a peak-finder algorithm. Once the peak frequencies and intensities are determined, they are sorted by intensity (from strongest to weakest). Next, the potential molecular carriers are determined for each peak by querying molecular spectroscopy databases and/or local (offline) catalogs for nearby transitions to the observed peak frequencies. For our application to rotational spectroscopy, we predominately queried the Splatalogue database \cite{splat} for all transitions within $0.5$\,MHz of the peak frequency. On average, there were approximately 10 transitions that were candidates for each observed line in the mixture spectra.

Once each molecular candidate is determined, a molecular embedder is used for the graph construction process. We have used the \textsc{mol2vec} model \cite{jae18} for our proof-of-concept work. This model was selected because the application in this work is focused on a mixture of small hydrocarbon molecules that are relevant to astrochemistry. Previously, several studies have shown that this \textsc{mol2vec} algorithm is effective for molecular embedding in astrochemical machine learning applications \cite{lee21}. That being said, this embedder has also been successfully applied in projects pertaining to various other fields, such as pharmaceutical sciences and drug candidate analysis \cite{dat21,ma21,che22,lee23}. Future work for our rotational spectroscopy application will explore the efficacy of more nuanced embedding models such as \texttt{VICGAE} \cite{orion}. Our technique is embedder-agnostic, however, and different approaches will certainly yield more optimal representations for additional applications.

\textsc{mol2vec} is an unsupervised machine learning method that re-purposes the \textsc{word2vec} algorithm \cite{word2vec} to create molecular feature vectors. Using a training set of molecules in the form of SMILES strings \cite{wei88}, the Morgan algorithm \cite{mor65} is first employed to create a dictionary of substructures within a certain radius of each atom. This dictionary is then fed into a multi-layer perceptron, which is trained to map each substructure to the surrounding substructures in every molecule. Consequently, similar vector representations are generated for substructures that appear in comparable chemical contexts. These substructure representations are subsequently summed to form molecular feature vectors. The \textsc{mol2vec} model that produced the molecular feature vectors for this work is the same as that described by Fried et al. (2023) \cite{fried23}. This model was trained on a dataset of 3,634,046 molecules collected from various online databases like Pubchem \cite{pubchem}, ZINC \cite{zinc}, and the NASA PAH database \cite{nasa2}. These molecules were tailored for astrochemical relevance, as they mostly contained less than 12 non-hydrogen atoms and were only composed of atoms in the first four rows of the periodic table. More expansive training sets, or those with molecules tailored for other applications, can certainly be used with our approach. 

With these molecular embeddings, the molecular graph required for the ranking algorithm can be constructed. The time required for the overall analysis heavily depends on the number of nodes in the graph. Therefore, the number of graph nodes is selected to balance the time required while ensuring that the graph is large enough to consider an ample number of molecular candidates. In this proof-of-concept demonstration, our graph consisted of a subset of 288,491 molecules from the 3.6 million molecules used to train the \textsc{mol2vec} model. The molecular nodes were then connected with bidirectional edges if the euclidean distance between their \textsc{mol2vec}-generated vector representations was less than an optimized threshold value. In its current construction, each node has on average $\sim$105 connections. 

To determine which hyperparameters, such as the distance threshold and distance metric, were critical to the graph generation and get an initial set of values for them, a 5-fold cross-validation grid search was run on the benzene/\ce{O2} discharge mixture dataset. The molecules in this mixture were separated into five splits. In each step, four of the five splits were inputted into the algorithm as ``detected" molecules. The hyperparameters were then determined as those that maximized the ranking of the molecules in the final split. Time efficiency was also considered when determining these parameters. 

\begin{figure}[h!]
\begin{center}
\includegraphics[width=\columnwidth]{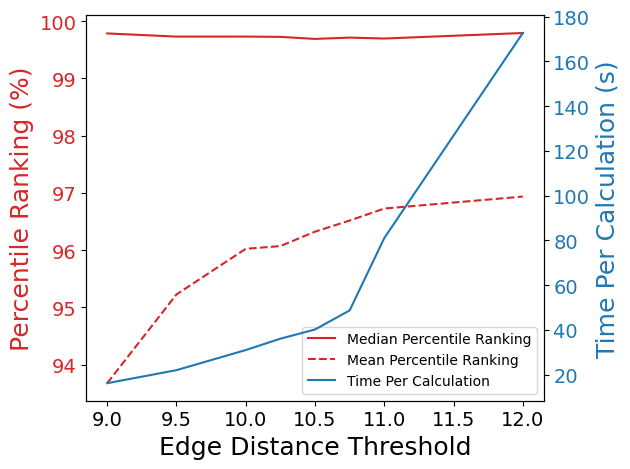}
\caption{Results from the 5-fold cross validation grid search for hyperparameter tuning. Only the results using a euclidean distance metric are shown. The individual values are listed in Table~\ref{table:hyperparams}. The x-axis displays the different edge connection distance thresholds that were tested for hyperparameter tuning. The red curves depict the median and mean percentile rankings of the molecules from the validations sets. The blue curve denotes the time required for each graph calculation. Additional details regarding the parameters and the cross-validation process are presented in the text.}\label{fig:hyperparams}
\end{center}
\end{figure}

The cross-validation grid search results for both a cosine and euclidean distance metric are displayed in Table~\ref{table:hyperparams}, and the  euclidean distance results are visualized in Figure~\ref{fig:hyperparams}. In the Table and Figure, the median and mean percentile rankings represent the percentile rankings of the molecules in each of the validation sets. For instance, a molecule with a structural relevance score higher than 99 percent of the molecules in the graph has a percentile ranking of 99\%. The time per calculation then reports how much time on average was required for the graph calculation to converge.

It can be seen that the performance is much stronger when euclidean distance is used as the distance metric as opposed to cosine distance. As shown in Figure~\ref{fig:hyperparams}, the median percentile ranking stays fairly consistent among the various distance thresholds. However, the mean scores improve notably as the number of edge connections increases. This suggests that even for a quick calculation in which the graph has few connections, the algorithm can effectively narrow-in on nearby chemically relevant molecules. However, with more edge connections, a broader range of the chemical space can be comprehensively explored. This, in turn, increases the score of relevant molecules that are further away from the region of chemical space where most ``detected" species are concentrated. This results in  fewer low-ranking ``outlier" molecules and increases the mean score, thereby improving the performance of the graph.

The optimal hyperparameters are a euclidean distance metric and a distance threshold of 10.75 for an edge to be placed between two nodes. This appears to be an inflection point in Figure~\ref{fig:hyperparams}, in which after this point the mean ranking percentile practically plateaus whereas the algorithm becomes much less efficient. 

\begin{figure}[h!]
\begin{center}
\includegraphics[width=\columnwidth]{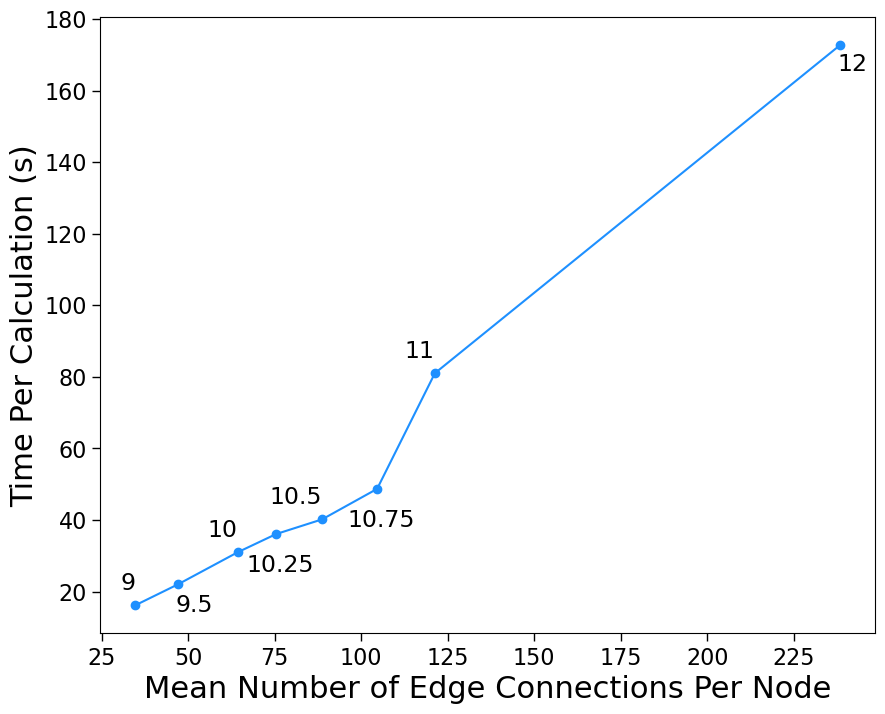}
\caption{Mean number of edge connections per node in the graph plotted versus the time required for the graph calculation to converge. The labels on the points depict the edge distance thresholds at each point. The individual values are listed in Table~\ref{table:hyperparams}.}\label{fig:hyperparam_times}
\end{center}
\end{figure}

The time required for the graph calculation to converge is also plotted against the average number of edge connections per node in Figure ~\ref{fig:hyperparam_times}. The time requirement increases fairly linearly from a distance threshold of 9 to 10.75. However, there is a notable time jump once the distance threshold reaches 11. This occurs because, for the hyperparameter tuning datasets, the graph requires an additional iteration to converge when the distance threshold reaches 11. In contrast, the number of calculation iterations remains consistent at all previous points. Therefore, the algorithm has to loop through all nodes and edges one extra time, thus resulting in the added time requirement.

\begin{table*}
\centering
\caption{5-fold cross validation grid-search results used for hyperparameter tuning. Details are presented in the text. The optimal hyperparameters were selected to be a euclidean distance metric with a distance threshold of 10.75 for an edge connection.}
\label{table:hyperparams}
\begin{tabular}{ccccccc}
\toprule
Distance& Distance & Mean \# & Mean Validation & Median Validation & Time Per \\
Metric & Threshold &Edge Connections & Percentile (\%) & Percentile (\%) & Calculation (s) \\
\midrule
Euclidean & 9 & 34.7 & 93.67 & 99.79 & 16.2\\
Euclidean & 9.5 & 47.0 & 95.23 & 99.73 & 22.0\\
Euclidean & 10 & 64.3 & 96.02 & 99.73 & 31.0\\
Euclidean & 10.25 & 75.4  & 96.07 & 99.73& 36.1\\
Euclidean & 10.5 & 88.7 & 96.32 & 99.69& 40.2  \\
Eudlidean & 10.75 & 104.6 & 96.52 & 99.71& 48.7\\
Euclidean & 11 & 121.3 & 96.72 & 99.70& 81.0\\
Euclidean & 12 & 238.4 & 96.94 & 99.79& 172.6\\
Cosine & 0.97 & 189.1 & 89.80 & 99.71 & 132.3\\
Cosine & 0.975 & 100.7& 87.50 & 99.61 & 47.6 \\
Cosine & 0.98 & 51.6 & 87.01 & 99.67 & 23.1\\

\bottomrule
\end{tabular}
\label{table:time}
\end{table*}

Ultimately, this hyperparameter tuning process required less than an hour of computation time and very minimal human intervention. These hyperparameters should be applicable to a plethora of other chemical mixtures, so long as the molecular contents are not wildly chemically dissimilar in future situations. For example, the hyperparameter set determined from the benzene/\ce{O2} discharge is probably sufficient for virtually any astrochemically relevant mixture, and likely for terrestrial mixtures, atmospheric, organic, and/or combustion mixtures. A mixture dominated by very large or inorganic species, on the other hand, would possibly necessitate a new graph generation and hyperparameter search.

Once the graph is constructed, the ranking algorithm is run. At the beginning of the process, some molecular priors can be provided to the algorithm as the ``detected" molecules. For example, the mixture spectra analyzed in this work were generated via the electrical discharge of several molecules. Therefore, these precursor molecules are initially entered as the ``detected" species. The ranking algorithm initializes all weight in the graph on the nodes corresponding to the ``detected" molecules. As an additional note, it is not strictly necessary to input known molecular priors for the mixture. Instead, for the first line, the assignment can rely solely on the spectroscopic match. The first molecule inputted into the graph as a ``detected" species would then be the assigned carrier of the strongest peak. In fact, for the mixtures that are tested in this work, the results stay almost fully consistent when beginning the assignment process with no molecular priors. This may be a potentially riskier approach, however, since a misassignment on the first line could result in the graph converging to the incorrect region of chemical space for the mixture. That being said, since the algorithm re-evaluates all previous assignments after each additional molecule is assigned, it will still typically recover from an incorrect initial assignment.

Once a detected molecule is inputted into the graph, the weight of each node is iteratively updated. In order to calculate the new node weights, we use the following equations:

\begin{equation}
R(p_i,t=0) =
\begin{cases}
  10 & \text{if $p_i \in$ detected} \\
  0 & \text{otherwise}
\end{cases}
\end{equation}

\begin{equation}
R(p_i, t+1) = R(p_i, t=0) + \sum_{p_j \in M(p_i)} \frac{R(p_j,t)}{1.5 \times L(p_j)}
\end{equation}

\begin{figure*}
\begin{center}
%\figurenum{1}

\includegraphics[width=\textwidth]{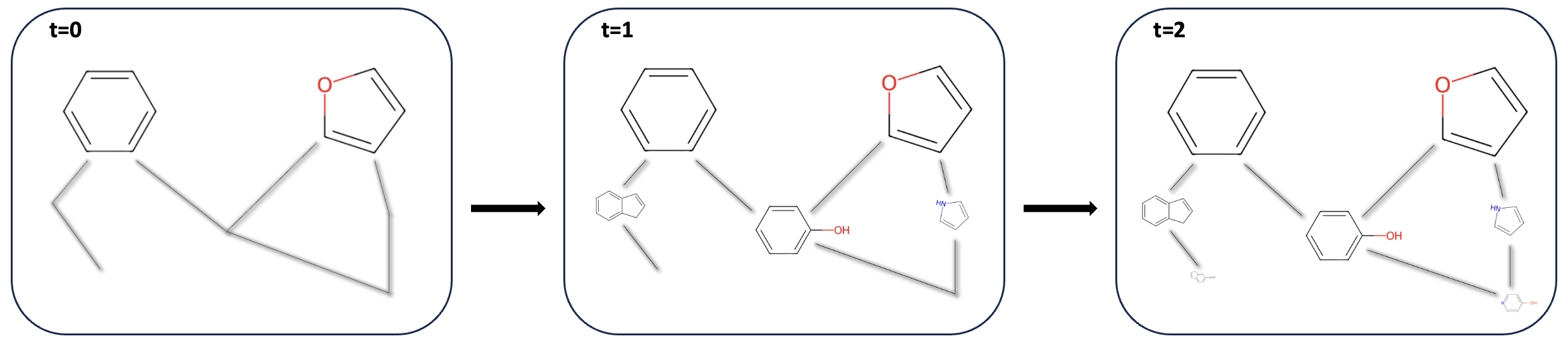}
\caption{Depiction of the weight of each node during several time-steps of the graph-based ranking system on a small molecular graph. The size of the molecules corresponds to the node weights. All weight is initialized on the detected molecules (benzene and furan in this example). During each time-step, the weight of every node is calculated based on the weights of the connected nodes. Therefore, at t=1, the molecules that are one connection away from the detected molecules gain weight, and at t=2 the molecules two connections away gain a reduced weight. Overall, through this process, molecules nearer to the detected species in chemical vector space become more highly ranked.}\label{fig:graph}
\end{center}
\end{figure*}

In these equations, $R(p_i,t)$ is the score of node $p_i$ at time-step $t$. $M(p_i)$ is the collection of all of the nodes that are connected via a bidirectional edge to node $p_i$. $L(p_j)$ is the number of nodes that are connected to node $p_j$. In summary, during every iteration, the weight of each node is updated based on the weight of its connected nodes. Therefore, in the first iteration, the nodes that are directly connected to the ``detected" molecules gain weight. The nodes that are two connections from the ``detected" species then gain a further reduced weight in the second time-step. This iterative process continues until the weight of each node converges (i.e. the weight of each node does not change by more than $10^{-5}$ between two iterations). The graph-based ranking system is depicted in Figure~\ref{fig:graph}. During each iteration, the weights of the detected molecules are also refreshed to their initial value, thus ensuring that the detected molecules are consistently the highest ranked in the graph. The division by $L(p_j)$ is included so as to limit any significant imbalances in the graph construction. For example, if there are certain nodes that have far more connections than most others, this division ensures that they will not have a disproportionate contribution to the ranking system. This overall process results in the nodes that are connected to several of the ``detected" molecules (and thus being close in chemical vector space to these species) becoming highly ranked.

After the ranking system converges, the resulting scores of the specific candidate molecules with a transition within $0.5$\,MHz from the strongest line in the spectrum are collected. Then the raw scores are converted into ranking percentiles based on the scores of all 288,491 nodes in the graph. 

Following the calculation and collection of the structural/chemical relevance scores, the algorithm proceeds to investigate how likely each candidate is based on the quantum mechanical properties of the molecular transitions in question. The following steps will differ based on the specific spectroscopy that is being used for the analysis. The next section will describe the steps taken for analyzing mixtures using rotational spectroscopy. The main considerations are the closeness of the frequency match along with the likelihood of the observed line intensity.

The frequency match is considered in a linear manner by multiplying the previously mentioned percentile scores by the following scaling factor. 

\begin{equation}
\text{Scaler} = 1 - \frac{|\nu_{\text{observed}} - \nu_{\text{catalog}}|}{5}
\end{equation}

The division by five in this equation was a hyperparameter determined in conjunction with other threshold hyperparameter values. This value was set so as to balance the requirement for a strong frequency match with the potentially large catalog frequency uncertainties.

Next, in order to investigate the line intensity likelihood, the spectra for each molecular candidate are simulated at $4\,$K, which is often the approximate rotational temperature reached in microwave spectroscopy experiments involving supersonic expansions. If available, the simulations can be forward modeled to include instrument response functions and other instrument-specific effects.  In this initial analysis, the instrument response is assumed to be flat. Spectral simulations are conducted using the \texttt{molsim} Python package \cite{molsim}. The following checks are then performed:

\begin{enumerate}
  \item If this line is the strongest occurrence of this molecule in the mixture spectrum, the transition should correspond to one of the strongest simulated transitions at the experimental temperatures. 
  \item The simulated catalog intensities are then scaled such that the catalog intensity of the line in question matches the observed line intensity in the mixture spectrum. There should not be any unreasonably strong predicted molecular transitions (i.e. much stronger than the most intense observed line in the mixture spectrum).
  \item If other lines of this molecule have been observed and assigned in the mixture, the relative strengths of the lines should be reasonable at the expected rotational temperatures.
  \item At least half of the predicted $10\sigma$ lines of the molecule should be present in the mixture spectrum. 
\end{enumerate}

If any of these conditions are violated, the score of the molecule is multiplied by a factor of $0.5$.  This scaling factor was chosen to sufficiently diminish the score of the molecule so that it is no longer in consideration for assignment. The current process of checking line intensities should work sufficiently for practically any experiment that investigates rotational signals. Additional work may still be required for spectra with drastically varying noise levels throughout the measured frequency range, however.

\begin{figure}[h!]
\begin{center}
\includegraphics[width=\columnwidth]{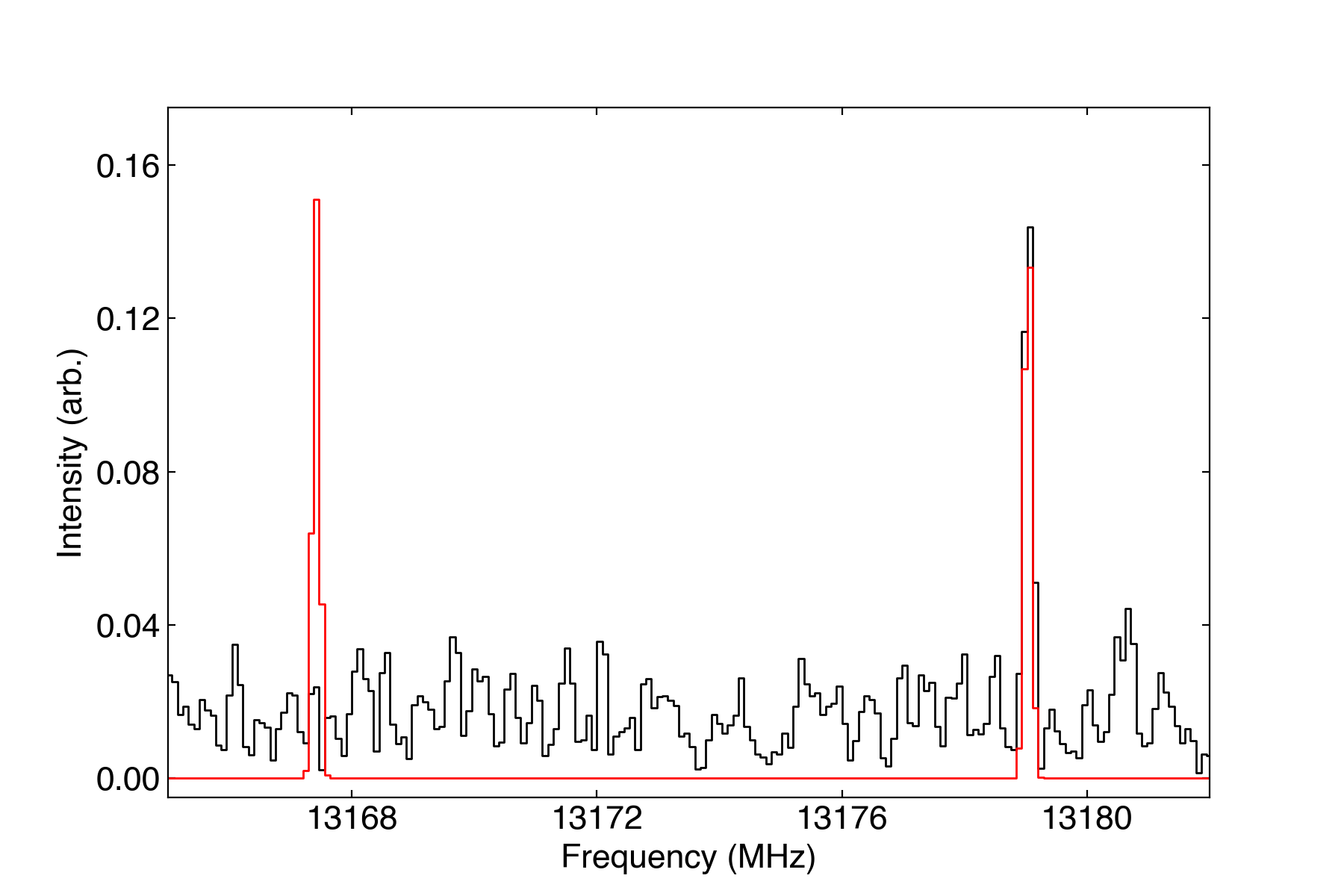}
\caption{Experimental rotational spectrum collected in the benzene discharge experiment (black) overlaid with the simulated spectrum of cyclohexadiene at 4\,K (red).  The simulated abundance of cyclohexadiene was scaled such that the intensity roughly matched the experimental peak at 13179 MHz. The absence of an experimental peak at 13167 MHz suggests that cyclohexadiene is not the molecular carrier of the 13179 MHz line.}\label{fig:intensity}
\end{center}
\end{figure}

The importance of investigating the line intensity is demonstrated in Figure~\ref{fig:intensity}. As can be seen, the line in the experimental spectrum (black) at 13179 MHz almost perfectly matches the frequency of a cyclohexadiene catalog transition (red). However, when that transition is simulated to match the intensity of the line in the experimental spectrum, a similar strength cyclohexadiene transition is expected to be seen around 13167 MHz. Since this line clearly does not appear in the experimental data, we have reason to rule out cyclohexadiene as the molecular carrier for the 13179 MHz line. 

The last considerations of the algorithm pertain to additional structural/chemical factors. Firstly, if the molecular candidate is isotopically substituted and the main isotopologue has not been observed in the spectrum or has been observed with line intensities that suggest unrealistic isotopic ratios, the molecular score is multiplied by a factor of $0.5$. Finally, a molecule can be further ruled out if it contains atoms that are unlikely to be present in the mixture. This is an adjustable parameter that can be inputted by the user. For example, for the datasets analyzed in this work, since the precursor molecules only contain carbon, hydrogen, nitrogen, and oxygen, the molecular scores of species containing other atoms (besides common contaminants such as sulfur) are multiplied by a factor of $0.5$. This can be tailored by the user because it can also depend on various environmental factors or contaminants from previous trials within an experimental setup. 

Following the numerous aforementioned considerations, each molecular candidate has a final likelihood score. In order to turn these raw scores into percentages for each molecule, a Softmax function is applied to the final scores. If both the raw final score and the Softmax percentage score of the top ranked molecule are above certain thresholds (93\% and 70\%, respectively), the algorithm then confidently assigns the transition to the top-ranked molecule. These threshold values were determined to be optimal for the datasets encountered in this work, however they can be easily adjusted for different applications if necessary. If the raw score threshold is not surpassed by the top ranked molecule, this indicates that the molecule is likely not the correct molecular carrier due to either a frequency, intensity, or structural relevance mismatch. In this case, the line is listed as ``unidentified." This label can either stem from the true molecular carrier not being present in spectroscopic databases or the structural/spectroscopic match not being convincing enough for confident assignment. 

%If several lines surpass the raw score threshold but not the Softmax percentage threshold, there is not enough evidence to rule-out either of these molecules. In these instances, the line is flagged for further analysis.  

If a molecule is confidently assigned, it is then added to the list of ``detected" molecules. Once this list is updated, either by the addition of a molecule or, as discussed momentarily, if one is removed following subsequent analysis, the graph is updated and new node weights are calculated now accounting for the additional (or removed) ``detected" species.

The same process as described above is then run on the next strongest observed line in the mixture spectrum. For this line, the structural relevance priors are more informed since they also include the molecule assigned to the strongest transition. Starting with the assignment of the second line and any updates to the graph, after each assignment the algorithm then proceeds to re-evaluate \emph{every} previously assigned line to ensure that the assigned molecular carrier stays consistent given the refined and more informed priors.

 Because calculating the graph weights is the slowest component of the algorithm, minimizing the number of graph calculations is crucial for efficiency. We found that for the mixtures encountered in this work, following the confident assignment of the first seven molecules, the graph properly converges to the correct regions of chemical space. Therefore, at this point, the graph weights only need to be updated again once five additional changes have been made to the detected molecules list. Therefore, in these mixtures containing upward of 50 unique molecular species, this drastically decreased the number of graph calculations required while maintaining equivalently strong results.

Finally, after each line in the mixture spectrum is investigated and assigned either to a molecular carrier or given an unidentified label, the structural relevance algorithm is run on the entire graph using the final list of ``detected" species. From this, we can identify which molecules in the graph are most structurally/chemically relevant to the mixture. These can then be starting points for further study (through calculation and/or experiment) either in real-time in parallel or in a follow-up investigation to assist in assigning the remaining unidentified features.

\section{Results}
\label{sec:results}
We used previously collected and analyzed mixture spectra from McCarthy et al. (2020) \cite{discharge} to test this algorithm. In this work, the authors combined chirped-pulse and cavity enhanced microwave spectroscopy to study the rotational spectrum of three mixtures produced by subjecting pure benzene as well as combinations of benzene with molecular oxygen and molecular nitrogen to an electrical discharge. This discharge fragments some precursor molecules, generates radicals and ions, and drives some populations into excited states, the combination of which results in a reaction mixture that produces a wide variety of product species. After measuring the broadband rotational spectrum of the mixture, the authors ultimately identified over 160 molecular products. Due to the large number of product molecules, the completeness and accuracy of the line assignment in these spectra, and the wealth of spectroscopic catalogs made available by this work, these mixture spectra were ideal candidates to test the algorithm.

Spectra of each mixture were collected in two frequency regions: $8-18\,$GHz and $18-26\,$GHz. The line intensities in these two regions differ notably, almost certainly due to differing response functions in the two microwave circuits used to perform the measurements. Since relative line intensities is a factor in the assignment algorithm, and because the instrument responses for these are not known, these frequency regions need to be analyzed separately. Our analysis presented in this paper focused on the $8-18\,$GHz portions of each spectrum. 

As mentioned previously, for each mixture, the initial ``detected" molecules inputted into the algorithm were the precursor species (i.e. benzene and molecular oxygen for the benzene/\ce{O2} discharge). Because there were three available mixtures, the benzene and benzene/\ce{O2} mixtures were used to determine the appropriate hyperparameters for the various algorithm components. The benzene/\ce{N2} mixture was used purely to validate the algorithm. 

The algorithm classifies spectral lines into the following categories: confidently assigned, more than one potential carrier, or unidentified. Lines that are confidently assigned have exactly one molecule for which the required calculation thresholds are met. Lines with multiple possible carriers have several molecules for which the raw score threshold is met. Unidentified lines have no molecules that meet these criteria.

\subsection*{Benzene Discharge}
\label{subsec:benzene_results}

In total, our dataset for the benzene discharge experiment contained 419 spectral lines assigned by McCarthy et al. (2020) \cite{discharge} from $6-20\,$GHz. We had spectroscopic catalogs for the molecular carriers of all but 13 lines. Therefore, our analysis mainly focused on the remaining 406 lines. Of these, the algorithm confidently assigned 384 (94.6\%) to a single molecular carrier. 373 of these agree with those determined by McCarthy et al. (2020) \cite{discharge}. For the lines for which there was a disagreeing assignment, we attempted to manually verify which molecule was correct. For these additional checks, we wanted to ensure that most other strong predicted transitions were present in the mixture spectrum and that the observed relative intensities were reasonable at the expected experimental temperatures. For 8 of the 11 disagreeing lines, we verified that the algorithm suggested a correct molecular assignment. For example, the algorithm assigned the three transitions at 8772 MHz, 12706 MHz and 16331 MHz to cyclohexa-2,4-dienone, while they were originally assigned to three distinct molecules. Nine additional transitions are also assigned to cyclohexa-2,4-dienone, and the aforementioned three all match the relative intensities reasonably well. In fact, these three transitions are some of the stronger predicted transitions of the molecule. Therefore, it is very likely that this molecule is accountable for these three lines. The lines from the benzene discharge for which the algorithm provided a different output than McCarthy et al. (2020) \cite{discharge} are listed in the Supporting Information (Table S1).  We provide a detailed, line-by-line assessment of the reasoning behind our assignment of the correct molecular carrier.

The algorithm labelled five lines (1.2\%) as having multiple possible molecular carriers. However, for four of these five lines, the algorithm suggested that the most likely carrier is the species assigned by McCarthy et al. (2020) \cite{discharge}. Finally, 17 lines were labelled as unidentified. As was done for the disagreeing confident assignments, we manually simulated the rotational spectra of the assigned molecules to verify why the algorithm labelled the line as unassigned. We ultimately agree with eight of these ``unidentified" assignments, since in these cases the assigned molecules have several simulated strong transitions that are not present in the experimental spectrum or were assigned based on an unrealistically weak predicted transition.

%\textbf{As can be seen in Table S1, for several of these algorithm misassignments, the issues are a result of the diminished instrument response around and above 18 GHz, since some of these molecules have strong transitions in this range.  Therefore, having a method to include information regarding the instrument response and intensity profile would likely be a very beneficial future addition to the algorithm.}

Overall, through this manual analysis, we only found a total of 12 clear misassignments from the algorithm. Thus, if we assume that the assignments for which McCarthy et al. (2020) \cite{discharge} and our algorithm agree are correct, the overall accuracy rate of the algorithm was $\sim$97.0\%. As an additional note, the line assignments were equivalent if no molecular priors were inputted into the algorithm.

\subsection*{Benzene/\ce{O2} Discharge}
\label{subsec:o2benzene_results}

For the discharge of benzene and molecular oxygen, our entire dataset contained 899 lines. We had the spectroscopic catalogs for the molecular carriers of 859 of these lines. Of these 859 lines, the algorithm provided confident assignments for 769 lines. The algorithm and McCarthy et al. (2020) \cite{discharge} provided the same molecular carrier for all but 34 of these 769 lines. For these 34 disagreements, we manually investigated each of the assigned molecules. Ultimately we are confident in 24 of the algorithm assignments, but found 10 clear misassignments. Thorough analysis of all disagreeing assignments from the benzene/\ce{O2} experiment are also listed in the Supporting Information (Table S2). 

29 lines were then assigned as having multiple possible molecular carriers. However, in 23 of these 29 instances, the highest ranked of these multiple possible carriers agreed with the assignment of McCarthy et al. (2020) \cite{discharge}.

Furthermore, the algorithm suggested that 61 of these 859 lines were ``unassigned," indicating that none of the molecular candidates were a convincing match. We manually investigated each of these lines. We ultimately found 11 clear misassignments by the algorithm. However, in the remaining cases, we agree with the algorithm's output. In most of these instances, the algorithm does not assign the molecular carrier because of issues with the relative intensity. Of note, the majority of these disagreeing assignments occur in the weakest 200 lines of the spectrum. This is fairly unsurprising since it is more difficult to compare relative line intensities when even the strongest lines are hardly above the noise level.

Overall, through the aforementioned manual analysis of the disagreeing lines, we only found 21 total obvious misassignments by the algorithm. Therefore, the overall accuracy rate was 838/859 ($97.6\%$). When running the assignment algorithm on this dataset with no inputted molecular precursors, only three lines ($0.35\%$) were assigned differently.

%\citet{discharge} derived rotational constants for several prominent unidentified molecules in the discharge experiments (Table 2 in their manuscript). Since , we proceeded to calculate the rotational constants of the 1,000 top ranked undetected molecules following the completion of the algorithm. This allowed us to investigate whether any of these molecules have rotational constants that match those determiend by \citet{discharge}. We used \texttt{PSI4} \citep{psi4} for these calculations. Due to the fairly large number of calculations, we used the low-cost B3LYP/6-31G(d) functional and basis set in order to increase efficiency. 

\subsection*{Benzene/\ce{N2} Discharge}
\label{subsec:n2benzene_results}
The prior two mixture analyses were used to tune our model.  As described earlier, however, once done, the algorithm should be applicable to similar mixture analyses with little to no tuning.  To test this, we used the third mixture presented in the work of McCarthy et al. (2020) \cite{discharge} and ran the algorithm ``out of the box."  For this mixture, our dataset contained a total of 717 lines that were identified by McCarthy et al. (2020) \cite{discharge}. Of these lines, we had spectroscopic catalogs for the carriers of all but 24, leaving 693 remaining transitions. Of these, the algorithm confidently assigned 638 ($\sim$92\%) to a single molecular carrier. In 585 of these instances, the algorithm assigned the same molecular carrier as McCarthy et al. (2020) \cite{discharge}. Following the same procedure as the previous two datasets, we manually investigated the 53 lines for which there was a disagreeing assignment through spectral simulation. Ultimately, of these 53 lines, we found only five clear misassignments by the algorithm. In the other instances the molecule suggested by the algorithm is a feasible match. However, there are also a few additional algorithm assignments for which complete certainty is challenging due to the weakness of some spectral lines. Thorough analysis of each of these lines is present in the Supporting Information (Table S3). 

For 24 of the transitions in this mixture, the algorithm suggested that there was more than one possible molecular carrier. However, in all but five instances, the highest ranked of these possible carriers agreed with the assignment of McCarthy et al. (2020) \cite{discharge}. Finally, the algorithm listed 31 of the 693 lines as ``unidentified." Through further manual analysis, we spotted only six clear misassignments from the algorithm.

Overall, in the dataset of 693 lines, we only identified 11 lines that were clearly misassigned by the algorithm, thus giving an accuracy rate of 98.4\%. We also ran the algorithm on this dataset with no molecular priors, and once again only three lines (0.43\%) were assigned differently.

As mentioned previously, this dataset was used solely for validation, while the others were utilized to tune the hyperparameters. It is therefore notable that the assignment accuracy on this dataset is very similar to the benzene and benzene/\ce{O2} experiments. This provides confidence that the current algorithm architecture can be successfully generalized to additional mixtures. 

Following the completion of the algorithm, the molecules in the graph with the highest structural relevance scores are strong candidates to be present in the mixture. Therefore, if experiments such as microwave double resonance can be used to determine unidentified lines that belong to a single molecular carrier and rotational constants are subsequently derived (as was done by McCarthy et al. (2020)\cite{discharge}), rotational constant calculations can be run on the top candidate molecules to check if these values match the experimentally determined constants. Certain quantum chemistry Python packages such as \texttt{Psi4} \cite{psi4} allow for the incorporation of these calculations into the algorithm. That being said, this process is complicated by the requirement for accurate and efficient conformational analysis. As mentioned previously, due to its structural sensitivity, different conformers of a molecule have completely unique rotational constants. Therefore, it is important that the calculation is conducted on the proper conformer of any given molecule. Certain additional Python packages, such as \texttt{Open Babel}\cite{openbabel}, can aid this process. 

As shown in Table~\ref{table:hyperparams} and Figure~\ref{fig:hyperparams}, the median percentile ranking of the molecules in the validation sets of the 5-fold cross validation was around 99.7\%.  In fact, of the 92 molecules in the combined validation sets, 24 of the molecules were ranked within the top 300 species in the respective graphs (out of just less than 300,000). Therefore, the graph-based ranking system shows a very strong ability to highly rank the molecules relevant to the mixture in question. This provides confidence that the molecular carriers of the unidentified lines would be prominently ranked. Thus, if one attempted to assign these unidentified lines by running electronic structure calculations on the structurally relevant molecules, it would possibly only take a reasonable number of fast calculations to find the correct carrier.

%As an example, there were a few molecules that were assigned by McCarthy et al. (2020) \cite{discharge} for which we did not have a spectroscopic catalog. Therefore, our algorithm was unable to assign these molecules and instead listed most of these lines as unidentified. In fact, hepta-1,2,4-trien-6-yne contributed to six of these lines. Following the completion of the algorithm, this molecule was ranked 171st out of all 288,491 molecules (99.94th percentile) in the network. Thus, if one attempted to assign these unidentified lines by running electronic structure calculations on the structurally relevant molecules, it would only take a reasonable number of fast calculations to find the correct carrier for this series of lines. 

\label{sec:results}

\section{Discussion}
\subsection{Analysis Speed}
The time taken for the algorithm to analyze the three datasets on a laptop (Apple M2 Pro chip, 16 GB memory, 10 core CPU) can be seen in Table~\ref{table:time}. This time requirement, ranging from 14 to 29 minutes, presents a several order of magnitude efficiency improvement over manual methods of assigning mixtures using broadband rotational spectroscopy. For example, the process of manually checking the 95 disagreeing assignments in the benzene/\ce{O2} discharge dataset required greater than 10 hours of work. In this process, only approximately $10\%$ of the lines in this dataset were investigated, and just a few of the molecular candidates were checked for each line. Thus, the entire process would have likely taken several weeks to complete through manual cross-referencing of spectroscopic catalogs. Notably, the current time requirement of the algorithm is also generally faster than the time-scales of these broadband rotational spectroscopy scans. Thus, this algorithm can possibly be used for real-time characterization of these spectra. 

\begin{table*}
\centering
\caption{Time required for the algorithm analysis on the three benzene discharge mixtures. }
\label{observed_lines}
\begin{tabular}{ccc}
\hline
Mixture & Number of Lines & Time Required (min)\\
\hline
Benzene & 419 & 14 \\
Benzene/\ce{N2} & 717 & 18 \\
Benzene/\ce{O2} & 899 & 29\\
\hline
\end{tabular}
\label{table:time}
\end{table*}

That being said, there are still avenues to increase the efficiency of the algorithm. For example, several aspects of the analysis, including the catalog scraping and the graph calculation could be made more efficient by greater parallelization within the code. Additionally, simply improving the code parallelization and construction would accelerate the analysis without reducing the algorithm's performance.

\subsection{Generalizability}
In this paper, our analysis of this structural relevance metric was limited to benzene-containing discharge experiments studied using rotational spectroscopy. As previously mentioned, the current graph model and hyperparameters should be usable ``out of the box" for mixtures that are relevant to atmospheric and astrochemical applications, as well as many terrestrial contexts, including combustion and organic mixtures. Additionally, the hyperparameters for the consideration of the spectroscopic match (i.e. frequency and relative intensity checks) should be applicable to almost any rotational spectroscopy experiment. However, further effort will be required to validate this method using more diverse mixtures and in conjunction with other spectroscopic techniques. This method is applicable to any analysis method that presents quantifiable signals which carry distinct molecular information and are accessible in a machine-readable catalog or database. In fact, the structural relevance component is completely separable from the quantum mechanical and spectroscopic analysis. In this case, the calculation of the structural relevance score is simply the first step in the assignment process and requires no insight from the spectroscopy. Therefore, the same code to calculate the structural relevance of a molecule can be directly applied to other techniques with practically no alteration. The only necessary changes would potentially be tweaking some of the model hyperparameters (i.e. distance threshold for an edge to be placed between two molecules and the molecular dataset used to generate the graph to better fit the dataset of the specific project).

In fact, this process could be even more useful in analysis techniques that have less separated signal features than those in rotational spectroscopy. Because of the structural specificity and high spectral resolution of rotational lines, there are typically a relatively small number of database transitions that are candidates for any given observed spectral peak. However, this is not always the case in other spectroscopic techniques. For example, in infrared (IR) spectroscopy, functional groups often present vibrational frequencies quite close to a canonical ``characteristic" frequency (e.g. the \ce{C=O} aldehyde stretch in formaldehyde is at 1746 cm$^{-1}$ and the same stretch in acetaldehyde is at 1743 cm$^{-1}$)\cite{nist}. Thus, molecules that are composed of similar functional groups can have heavily overlapping peaks in a mixture spectrum. Additionally, in mass spectrometry (MS), molecular fragments that are produced are indistinguishable if they have the same mass and charge. Therefore, molecules that have the same mass and fragment into similar sets of molecular ions can have nearly indistinguishable mass spectra. For example, Song et al. (2021) \cite{song21} showed that the 10 strongest fragment ions produced in collision-induced dissociation (CID)-MS/MS experiments of the isomers glucose and inositol are practically indistinguishable. Therefore, without employing more advanced experimental methods (also detailed in the work of Song et al. (2021) \cite{song21}), these spectra are challenging to differentiate. In these cases, a structural relevance metric that can compute how likely a molecular structure is given certain priors or other mixture components would be an extremely useful tool to determine which molecule is the true carrier of the observed spectral signal.

\section{Conclusion}
Due to the overlapping of the signals produced from spectroscopic methods, the determination of chemical mixture components using these techniques can be a challenging problem. That being said, real-world chemical mixtures are oftentimes composed of molecules that are structurally and/or chemically related to one another. Therefore, we devised a methodology for incorporating the structural/chemical relevance of a molecule in the analysis of mixture components. To do this, we employed a graph-based architecture in which the graph connectivity was dependent on relationships between molecular feature vectors that were produced using machine-learning embedding methods. Through an iterative process, this graph provided a score to each molecule that quantified how similar it is to the other identified mixture components or certain chemical priors. 

To demonstrate the efficacy of this structural relevance metric, we incorporated it into a rotational spectroscopy analysis algorithm. This was used to assign the molecules present in several mixtures that were originally measured and analyzed by McCarthy et al. (2020) \cite{discharge}. Ultimately, the algorithm provided extremely accurate molecular assignments ($\geq97\%$ in each of the mixtures analyzed). As currently constructed, this method is already much more efficient than manual analysis while requiring practically no human intervention and maintaining very high levels of accuracy. Thus, it presents a notable step-forward in automated mixture analysis methods.

\begin{suppinfo}

\textbf{Tables S1-S3:} Three long tables that describe the lines from all three discharge mixtures for which the AMASE algorithm and McCarthy et al.  (2020) \cite{discharge} assign different molecular carriers (PDF).

\textbf{Data S1:} Dataset for the benzene discharge mixture containing all molecular candidates for the transitions in the mixture (CSV).

\textbf{Data S2:} Text file describing each line assignment in the benzene discharge mixture (TXT). 

\textbf{Data S3:} Excel file containing all of the line assignments in the benzene discharge mixture. (CSV).

\textbf{Data S4:} Dataset for the benzene/\ce{O2} discharge mixture containing all molecular candidates for the transitions in the mixture (CSV).

\textbf{Data S5:} Text file describing each line assignment in the benzene/\ce{O2} discharge mixture (TXT). 

\textbf{Data S6:} Excel file containing all of the line assignments in the benzene/\ce{O2} discharge mixture. (CSV).

\textbf{Data S7:} Dataset for the benzene/\ce{N2} discharge mixture containing all molecular candidates for the transitions in the mixture (CSV).

\textbf{Data S8:} Text file describing each line assignment in the benzene/\ce{N2} discharge mixture (CSV). 

\textbf{Data S9:} Excel file containing all of the line assignments in the benzene/\ce{N2} discharge mixture. (CSV).

\textbf{Interactive Outputs:} HTML files that contain interactive outputs for the  discharge mixture assignments can be found in the following Zenodo repository: \url{https://zenodo.org/records/13380064} 

\textbf{Code File:} Python scripts to run the assignment algorithm on a mixture studied with rotational spectroscopy along with the hyperparameter tuning are available in the following GitHub repository: \url{https://github.com/zfried/AMASE}

\end{suppinfo}

\begin{acknowledgement}

We would like to sincerely thank the two anonymous referees who each provided expert criticisms which substantially improved the quality of this manuscript. The authors gratefully acknowledge the support of Schmidt Family Futures. The authors also thank Dr. Michael C. McCarthy for helpful discussion regarding the mixtures studied in this work and the previous assignment approaches. 

\end{acknowledgement}

\providecommand{\latin}[1]{#1}
\makeatletter
\providecommand{\doi}
  {\begingroup\let\do\@makeother\dospecials
  \catcode`\{=1 \catcode`\}=2 \doi@aux}
\providecommand{\doi@aux}[1]{\endgroup\texttt{#1}}
\makeatother
\providecommand*\mcitethebibliography{\thebibliography}
\csname @ifundefined\endcsname{endmcitethebibliography}  {\let\endmcitethebibliography\endthebibliography}{}

%\bibliography{achemso-demo}

\end{document}